# Large magnetoelectric coupling in multiferroic oxide heterostructures assembled via epitaxial lift-off

D. Pesquera[1,*], E. Khestanova[2], M. Ghidini[3,4,1], S. Zhang[1,5], A. P. Rooney[6], F. Maccherozzi[4], P. Riego[1,7,8], S. Farokhipoor[9], J. Kim[1], X. Moya[1], M. E. Vickers[1], N. A. Stelmashenko[1], S. J. Haigh[6], S. S. Dhesi[4] and N. D. Mathur[1,†]

[1]Department of Materials Science, University of Cambridge, Cambridge, CB3 0FS, UK
[2]ITMO University, Saint Petersburg 197101, Russia
[3]Department of Mathematics, Physics and Computer Science, University of Parma, 43124 Parma, Italy
[4]Diamond Light Source, Chilton, Didcot, Oxfordshire, OX11 0DE, UK
[5]College of Science, National University of Defense Technology, Changsha 410073, China
[6]School of Materials, University of Manchester, Manchester M13 9PL, UK
[7]CIC nanoGUNE, E-20018 Donostia-San Sebastian, Spain
[8]Department of Condensed Matter Physics, University of the Basque Country, UPV/EHU, E-48080 Bilbao, Spain
[9]Zernike Institute for Advanced Materials, University of Groningen, 9747 AG Groningen, The Netherlands

*dpesquera@cantab.net, †ndm12@cam.ac.uk

**The strain-dependent functional properties of epitaxial transition-metal oxide films can be significantly modified via substrate selection. However, large lattice mismatches preclude dislocation-free epitaxial growth on ferroelectric substrates, whose strain states are modified by applied electric fields. Here we overcome this mismatch-problem by depositing an epitaxial film of ferromagnetic $La_{0.7}Sr_{0.3}MnO_3$ on a single-crystal substrate of well-lattice-matched $SrTiO_3$ via a film of $SrRuO_3$ that we subsequently dissolved, permitting the transfer of unstrained $La_{0.7}Sr_{0.3}MnO_3$ to a ferroelectric substrate of $0.68Pb(Mg_{1/3}Nb_{2/3})O_3$-$0.32PbTiO_3$ in a different crystallographic orientation. Ferroelectric domain switching, and a concomitant ferroelectric phase transition, produced large non-volatile changes of magnetization that were mediated by magnetic domain rotations at locations defined by the microstructure — as revealed via**

**high-resolution vector maps of magnetization constructed from photoemission electron microscopy data, with contrast from x-ray magnetic circular dichroism. In future, our method may be exploited to control functional properties in dislocation-free epitaxial films of any composition.**

Transition-metal-based perovskite oxides display a wide range of functional properties that are relevant for current and future applications[1,2] including data storage[3,4], thin-film transistors[5], cooling[6], and the conversion of thermal[7,8], mechanical[9], solar[10] and chemical[11,12] energy with respect to electrical energy. Given that these functional properties are governed by spin, charge and orbital degrees of freedom that depend on interatomic distances and bond angles[13,14], the strain that exists in epitaxial films on single-crystal oxide substrates can promote metallicity[15], create ferroelectricity[16], modify magnetic ground states[17,18], tune charge order[19] and enhance catalytic activity[20,21]. Changing the composition and crystallographic orientation of the substrate therefore permits a chemically unique film to display different functional properties, expanding the scope for applications. However, accessible strain states are limited by a discrete set of lattice parameters, and may not possess equivalent microstructures. It is therefore attractive to use large voltage-driven strain from ferroelectric oxides to continuously tune the lattice parameters of epitaxial films, most notably for the electrical control of magnetization in ferromagnetic oxides[22,23], as proposed for electric-write magnetic-read data-storage cells[24–26].

Unfortunately, epitaxial films grown on ferroelectric substrates experience high degrees of strain and large dislocation densities as a consequence of lattice-parameter mismatches[20,21,25-27] that typically exceed >3%. Such large mismatches can only be partially mitigated via an epitaxial buffer layer[30], and the orientation of the voltage-driven strain is necessarily constrained by the epitaxy. Although one may instead grow an epitaxial film on a well-lattice-matched substrate that is subsequently thinned and glued to a ferroelectric substrate in any orienation[31–33], strain transmission is compromised by the inactive growth substrate and the glue, such that we observed no magnetoelectric effects when trying to couple strain-sensitive[22,34] epitaxial films of ferromagnetic $La_{0.7}Sr_{0.3}MnO_3$ (LSMO) to piezoelectric substrates[35] of $0.68Pb(Mg_{1/3}Nb_{2/3})O_3$-$0.32PbTiO_3$ (PMN-PT) via growth substrates of $SrTiO_3$ (STO) that had been thinned to 50-100 μm.

Here we exploit the afore-mentioned ferroic materials to investigate strain-mediated magnetoelectric effects after epitaxially growing LSMO/SRO//STO (001), dissolving the SrRuO$_3$ (SRO), and transferring the LSMO (001)$_{pc}$ film to a substrate of platinized PMN-PT (011)$_{pc}$ whose surface is known to display electrically driven changes of strain that are non-volatile[35] (pc denotes pseudocubic). This method of transfer avoids the epitaxial strain that would arise from the large ~3% lattice mismatch between LSMO and PMN-PT, while the ~1% mismatch between LSMO, SRO and STO is considerably smaller. The transfer of oxide films to ferroelectric substrates has not been hitherto reported, even though oxide films have been transferred to silicon substrates after likewise dissolving sacrificial layers[36,37]; MoS$_2$ trilayers have been transferred to ferroelectric substrates after dissolving growth substrates[38]; oxide films have been transferred to non-ferroelectric substrates after dissolving growth substrates[39,40]; and epitaxial oxide flakes[39] and MoS$_2$ monolayers[40] have been transferred to flexible substrates.

Ferroelectric domain switching in our PMN-PT substrate was accompanied by a rhombohedral-orthorhombic phase transition, and the resulting piezostrain led to an electrically switchable two-fold magnetic anisotropy in the transferred LSMO film, yielding magnetoelectric effects that were large, non-volatile and repeatable. Cross-sectional scanning transmission electron microscopy (STEM) revealed that these magnetoelectric effects were mediated by a serendipitously formed interfacial layer, based primarily on amorphous SiO$_x$. Photoemission electron microscopy (PEEM) with contrast from x-ray magnetic circular dichroism (XMCD) was used to obtained magnetic vector maps, which revealed the existence of micron-sized magnetic domains with invariant perimeters. These domains underwent electrically driven magnetic rotations through a range of different angles, implying the imprint of inhomogeneous stress during transfer.

**Results**

**Sample fabrication and characterisation**. An elastomer membrane of polydimethylsiloxane (PDMS) was used to transfer a 45 nm-thick layer of LSMO from its STO (001) growth substrate to platinized PMN-PT (011)$_{pc}$, after dissolving the intervening epitaxial layer of 30 nm-thick SRO with NaIO$_4$ $_{(aq)}$ (Fig. 1a). The $a \parallel [100]_{pc}$ and $b \parallel [010]_{pc}$ axes of LSMO that lay parallel to the film edges were closely aligned with the $x \parallel [100]_{pc}$ and $y \parallel [01\bar{1}]_{pc}$ axes of PMN-PT that lay parallel to the edges of the slightly larger substrate (Supplementary Note 1).

For simplicity, samples will be labelled LSMO:PMN-PT, without reference to the Pt electrodes on either side of the PMN-PT substrate, and without reference to the flat amorphous interfacial layer (Fig. 1b). STEM chemical analysis revealed this interfacial layer to be primarily composed of Si, O and C (Supplementary Note 2), implying partial degradation of the PDMS membrane during the SRO etch. Two other polymers[41] that we used for transfer retained only small LSMO flakes after etching.

X-ray diffraction (XRD) measurements of our LSMO/SRO//STO (001) precursor (red data, Fig. 1c) confirmed that the LSMO layer experienced tensile in-plane epitaxial strain. The high quality of the LSMO film was confirmed by the presence of thickness fringes, and a narrow $002_{pc}$ rocking curve of full-width-half-maximum 0.2° (red data, inset of Fig. 1c). Moreover, XRD reciprocal space maps around the STO 103 reflection (not shown) confirmed a good match between all three in-plane lattice parameters. XRD measurements of the LSMO film after it had been transferred to the platinized PMN-PT $(011)_{pc}$ substrate revealed that the epitaxial strain had been completely released (blue data, Fig. 1c), and that the full-width-half-maximum of the $002_{pc}$ rocking curve had increased to 1.7° (blue data, inset of Fig. 1c). This enhancement of texture is attributed to the surface topography of the platinized PMN-PT $(011)_{pc}$ substrate (Supplementary Note 3), whereas cracking was observed in the LSMO film both before and after transfer (Supplementary Note 4).

The release of epitaxial strain increased the LSMO saturation magnetization by 19%, from $2.27 \pm 0.03$ $\mu_B$/Mn after growth to $2.7 \pm 0.1$ $\mu_B$/Mn after transfer (Fig. 1d), because double exchange was enhanced in bonds that had been shortened by strain release[42]. The release of epitaxial strain also modified the biaxial magnetic anisotropy of the LSMO film (Fig. 1e). After growth, the in-plane LSMO $<110>_{pc}$ directions were magnetically easy due to magnetoelastic anisotropy arising from the biaxial in-plane strain imposed by the STO substrate[43–45]; after strain release and film transfer, the in-plane LSMO $<100>_{pc}$ directions were magnetically easy due to uniaxial magnetocrystalline anisotropy in each twin variant of the now twinned film[46] (Supplementary Note 5). Given that the transfer process reduces the magnitude of the easy-axis anisotropy (Fig. 1e), the increase of easy-axis coercivity (Fig. 1d) is inferred to arise extrinsically from the enhancement of microstructural complexity.

**Electrically driven strain in PMN-PT.** After thermally depolarizing PMN-PT in order to set zero strain, a bipolar cycle of electric field $E$ produced orthogonal in-plane strains $\varepsilon_x$ and $\varepsilon_y$ that took opposite signs to each other at almost every field, and underwent sign reversal near the coercive field (solid butterfly curves, Fig. 2a). Given that the two butterfly curves would be interchanged[35] if they arose purely from ferroelectric domain switching in rhombohedral PMN-PT, we infer that polarization reversal was instead associated with a phase transition[39-41], as confirmed by measuring XRD reciprocal space maps while applying an electric field (Supplementary Note 6). Large fields promoted the orthorhombic (O) phase by aligning the polarization along an out-of-plane <011>$_{pc}$ direction, whereas switching through the coercive field promoted rhombohedral (R) twins whose polarizations lay along the subset of <111>$_{pc}$ directions with an out-of-plane component. (A similar argument would hold if this latter phase were monoclinic[50] rather than rhombohedral.)

A minor electrical loop (blue dots in Fig. 2a) permitted two strain states to be created in PMN-PT at electrical remanence[35,47], with $\varepsilon_x = -0.16\%$ in state A, and $\varepsilon_x = +0.02\%$ in state B. The corresponding reciprocal space maps obtained at zero electric field (Fig. 2b-e) show that the O phase dominated in state A (single 222$_{pc}$ reflection, split 031$_{pc}$ reflection), while the R phase dominated in state B (split 222$_{pc}$ reflection, single 031$_{pc}$ reflection). The resulting structural changes in the LSMO film could also be detected by XRD (Supplementary Note 5), despite the twinning and the topography of the underlying ferroelectric domains.

**Strain-mediated electrical control of macroscopic magnetization in LSMO:PMN-PT.**
The biaxial magnetic anisotropy that we observed after transfer (Fig. 1e) was rendered uniaxial at both A and B during the course of 30 bipolar electrical cycles (Supplementary Note 7). Subsequent bipolar cycles modified the *x* and *y* components of magnetization by ~100% (solid butterfly curves, Fig. 3), and similar results were observed for two similar samples (Supplementary Note 8). The peak magnetoelectric coefficient of $\alpha = \mu_0 dM_x/dE = 6.4 \times 10^{-8}$ s m$^{-1}$ is similar to the value reported[23] for an LSMO film that benefited from good epitaxial coupling with a PMN-PT (001) substrate. The interconversion of remanent states A and B (blue dots in Fig. 3) rotated the single magnetic easy axis by 90° (Fig. 4), while the finite loop-squareness minimum in state B (Fig. 4d) may represent a vestige of the original four-fold anisotropy, or uniaxial regions trapped from state A.

**Strain-mediated electrical control of microscopic magnetization in LSMO:PMN-PT.**
The local magnetization of the thermally demagnetized film was imaged using photoemission electron microscopy (PEEM) with contrast from x-ray magnetic circular dichroism (XMCD). The resulting vector maps of the in-plane magnetization direction $\phi$ revealed that the electrically remanent A and B states were magnetically inhomogeneous within a 20 μm field of view (Fig. 5a,b); and that switching from A to B rotated the net magnetization in our limited field of view towards the $x$ axis (Fig. 5c,d), consistent with our macroscopic measurements of magnetic anisotropy (Fig. 4c,d). The magnetization was reasonably homogeneous within micron-sized domains whose perimeters coincided partly with cracks (Supplementary Note 9), and the electrically driven magnetic rotations in 1.3 μm-diameter regions (1-3 in Fig. 5a,b,e) were found to range from large (64° in region 1) to medium (-36° in region 2) to small (16° in region 3) (Fig. 5f). A magnetic free energy model approximately reproduces both the local and macroscopic magnetoelectric effects after including a spatially varying uniaxial magnetic anisotropy to mimic stress in the transferred film (Supplementary Note 10).

**Discussion**

Our experimental work may be summarized as follows. We first released an epitaxial film of LSMO from a single-crystal substrate of STO by etching an intervening epitaxial film of SRO. We then used a flexible elastomer membrane to transfer the released film of LSMO to a substrate of PMN-PT. Conformal contact with the surface topography of micron-sized ferroelectric domains yielded structurally contiguous single magnetic domains of strain-released LSMO. These magnetic domains typically spanned a few microns, such that they are larger than the majority of epitaxial magnetic structures defined by lithography[51–53].

The magnetoelectric effects that we observed both macroscopically and microscopically evidence good strain-mediated coupling. Therefore our result could be exploited in order to electrically control the free layer of a magnetic tunnel junction for data storage[24–26]. The release of epitaxial strain would avoid the performance-limiting suppression of the Curie temperature[54], and the free choice of in-plane misorientation angle would permit the realization of schemes for electrically driven magnetization reversal[55–57]. More generally, the physical and chemical properties of any epitaxial film could be electrically controlled after transfer to a ferroelectric substrate, whose composition and orientation may be freely selected. For example, one could in future achieve electrically driven strain-mediated inversion of orbital occupancy, thus permitting surface reactivity to be modified *in situ* rather than via changes of stoichiometry[58,59] or substrate[21].

**Acknowledgements**

D. P. acknowledges Agència de Gestió d'Ajuts Universitaris i de Recerca (AGAUR) from the Catalan government for Beatriu de Pinós postdoctoral fellowship (2014 BP-A 00079). E. K. acknowledges support by the Ministry of Science and Higher Education of Russian Federation, goszadanie no. 2019-1246. X. M. acknowledges funding from the Royal Society. We thank Diamond Light Source for time on beamline I06 (proposal SI14745-1). S. J. H. and A. P. R. acknowledge funding from EPSRC (Grant EP/P009050/1, EP/M010619/1 and the NoWNano DTC) and the European Research Council (ERC) (ERC-2016-STG-EvoluTEM-715502 and ERC Synergy HETERO2D). P. R. acknowledges funding by the "la Caixa" Foundation (ID 100010434). We thank Jiamian Hu, Manish Chhowalla, Emilio Artacho, Paul Attfield, Sohini Kar-Narayan and Judith Driscoll for discussions.

**Methods**

**Samples.** We fabricated three similar LSMO:PMN-PT samples (A, B and C). The LSMO film (edges along $a \parallel [100]_{pc}$ and $b \parallel [010]_{pc}$) was misaligned with the PMN-PT substrate (edges along $x \parallel [100]_{pc}$ and $y \parallel [01\bar{1}]_{pc}$) by 5° (samples A, B) and 20° (sample C). All experimental data were obtained using sample A or its precursor components, with the following exceptions: measurements of strain and electrical polarization were obtained using PMN-PT from the same master substrate that we used for sample A; AFM data were obtained using sample B; STEM data were obtained using sample C; and the macroscopic magnetoelectric data in Supplementary Fig. 7 were obtained using samples A-C.

**Epitaxial growth of LSMO/SRO bilayers.** Epitaxial LSMO (45 nm)/SRO (30 nm) bilayers were grown by pulsed laser deposition (KrF excimer laser, 248 nm, 1 Hz) on STO (001) substrates (5 mm × 5 mm × 1 mm) that had been annealed in flowing oxygen for 90 minutes at 950 °C. The SRO was grown in 10 Pa $O_2$ at 600 °C (1200 pulses, 1.5 J cm$^{-2}$). The LSMO was grown in 15 Pa $O_2$ at 760 °C (1800 pulses, 2 J cm$^{-2}$). After growth, the LSMO/SRO//STO stacks underwent *in situ* annealing in 50 kPa $O_2$ for 1 hour at 700 °C. Using x-ray reflectivity measurements, the growth rate for single layers of SRO and LSMO were both found to be ~0.025 nm/pulse.

**Platinized PMN-PT substrates.** Each $0.68Pb(Mg_{1/3}Nb_{2/3})O_3$-$0.32PbTiO_3$ (011)$_{pc}$ substrate (PMN-PT, Atom Optics) was cut to ~5 mm × ~5 mm × 0.3 mm from a different 10 mm × 10 mm × 0.3 mm master. Sputter deposition of Pt resulted in a 6 nm-thick top electrode (that served as ground), and a much thicker bottom electrode.

**Transfer of LSMO.** PDMS stamps were cut to 5 mm × 5 mm × 1.5 mm from a commercial specimen (Gelfilm from Gelpak), and each was brought into conformal contact with a given LSMO/SRO//STO stack by heating in air at 70 °C for 10 minutes (conformal contact was verified by the change in optical reflectance on elimination of the air gap). After floating the resulting PDMS/LSMO/SRO//STO stacks in $NaIO_{4\,(aq)}$ (0.4 M) for several hours, the SRO layers dissolved to release bilayers of PDMS/LSMO, which were washed with deionized water, and dried with $N_2$ gas. Using tweezers, each PDMS/LSMO bilayer was subsequently transferred to a platinized PMN-PT substrate, which had been previously cleaned using acetone and isopropanol, and recently cleaned by annealing in air at 120 °C for 10 minutes. After transfer, the entire stack was annealed in air (at 100°C for 10 minutes) to promote adhesion at the newly formed interface. After cooling to 70 °C and peeling off the PDMS stamp with tweezers, interfacial adhesion was further improved by annealing in air at 150 °C for 10 minutes.

**X-ray diffraction.** We acquired $2\theta$-$\omega$ scans and rocking curves for LSMO with a Panalytical Empyrean diffractometer (Cu-K$\alpha_1$, 1.540598 Å), using a hybrid 2-bounce primary monochromator on the incident beam, and a two-bounce analyser crystal before the proportional point detector. Reciprocal space maps of PMN-PT were acquired with the same incident beam optics and a PIXcel$^{3D}$ position-sensitive detector, using the frame-based 1D mode with a step time of 10 s.

We used Sample A and its epitaxial precursor to obtain $2\theta$-$\omega$ scans and rocking curves before applying an electric field (Fig. 1c). Our electric-field-dependent XRD data were also obtained using sample A, after acquiring a subset of the magnetoelectric data and then repeating the last anneal of the fabrication process (10 minutes in air at 150 °C) in order to depolarize the substrate. We first obtained reciprocal space maps of PMN-PT at successively larger positive fields after negative poling (Supplementary Fig. 5d,e), before acquiring reciprocal space maps for remanent states A and B (Fig. 2b-d). We then obtained $2\theta$-$\omega$ scans of LSMO for remanent states A and B (Supplementary Fig. 6).

**Atomic force microscopy.** Atomic force microscopy (AFM) images were obtained in tapping mode using a Veeco Digital Instruments Dimension D3100 microscope.

**Electron microscopy.** Cross-sectional transmission electron microscopy (TEM) specimens were prepared via an *in situ* lift-out procedure in a dual-beam instrument (FEI Nova 600i) that incorporated a focused ion beam microscope and scanning electron microscope in the same chamber. Both 5 kV and 2 kV ions were used to polish the TEM lamella to a thickness of 50 nm, and remove side damage. High-resolution scanning transmission electron microscopy (STEM) was performed using a probe-side aberration-corrected FEI Titan G2, operated at 80-200 kV with a high-brightness field-emission gun (X-FEG). Bright-field STEM imaging was performed using a probe convergence angle of 21 mrad and a probe current of ~90 pA. In bright-field images, identification of each atomic layer was achieved by elemental analysis using energy dispersive X-ray (EDX) and electron energy loss spectroscopy (EELS). EDX images were obtained using a Super-X four silicon drift EDX detector system with a total-collection solid angle of 0.7 srad. EELS images were obtained using a Gatan Imaging Filter (GIF) Quantum ER system, with an entrance aperture of 5 mm. The lamella was oriented by using the Kikuchi bands to direct the electron beam down the $[01\bar{1}]_{pc}$ zone axis of PMN-PT.

**Strain measurements.** Platinized PMN-PT (derived from the master substrate used for sample A) was cleaned like sample A, using acetone and isopropanol, and then annealed in air at 150 °C for 30 minutes in order to mimic the final depolarizing heat treatment experienced by sample A. A biaxial strain gauge (KFG-1-120-D16-16 L1M3S, Kyowa) was affixed using glue (CC-33A strain gauge cement, Kyowa) to the top electrode, with measurement axes along *x* and *y*. The initial values of resistance were used to identify zero strain along the two measurement directions. Strain-field data were obtained while applying bipolar triangular voltages at 0.01 Hz in the range ±10 kV cm$^{-1}$.

**Macroscopic magnetization measurements.** These were performed using a Princeton Measurements Corporation vibrating sample magnetometer (VSM), with electrical access to the sample as shown in ref.[22]. All data are presented after subtracting the diamagnetic contribution of substrate.

**Magnetic vector maps.** After completing all macroscopic magnetoelectric measurements, we obtained raw images of sample A after thermal demagnetization. The electrically remanent states A and B were interconverted *in situ* using a 300 V power supply that was connected via feedthroughs in the sample holder.

Data were obtained on beamline I06 at Diamond Light Source, where we used an Elmitec SPELEEM-III microscope to map secondary-electron emission arising from circularly polarized x-rays that were incident on the sample surface at a grazing angle of 16°. The probe depth was ~7 nm, and the 20 μm-diameter field of view that we used in this study implies a lateral resolution of ~50 nm, with each pixel representing ~20 nm.

Raw images were acquired during 1 s exposure times with right (R) and left (L) circularly polarized light, both on the Mn $L_3$ resonance at 645.5 eV, and off this resonance at 642 eV. The pixels in a raw XMCD-PEEM image describe the XMCD asymmetry $(I^R - I^L)/(I^R + I^L)$, which represents the projection of the local surface magnetization on the incident-beam direction. Here, $I^{R/L} = (I_{on}^{R/L} - I_{off}^{R/L})/I_{off}^{R/L}$ denotes the relative intensity for secondary electron emission due to x-ray absorption on ($I_{on}^{R/L}$) and off ($I_{off}^{R/L}$) the Mn $L_3$ resonance (the comparison between intensities obtained on and off resonance avoids the influence of any inhomogeneous illumination).

We averaged 40 raw XMCD-PEEM images to obtain a single XMCD-PEEM image for each of two orthogonal sample orientations. These two images were combined in order to yield vector maps of in-plane magnetization, after correcting for drift and distortion via an affine transformation that was based on topographical images of x-ray absorption for each sample orientation. Each of these topographical images was obtained by averaging all raw images that had been obtained on resonance with left and right-polarized light.

**XAS images.** X-ray absorption spectroscopy images are presented alongside XMCD-PEEM images (Supplementary Note 9) by plotting $I^R + I^L$ and $(I^R - I^L)/(I^R + I^L)$, respectively.

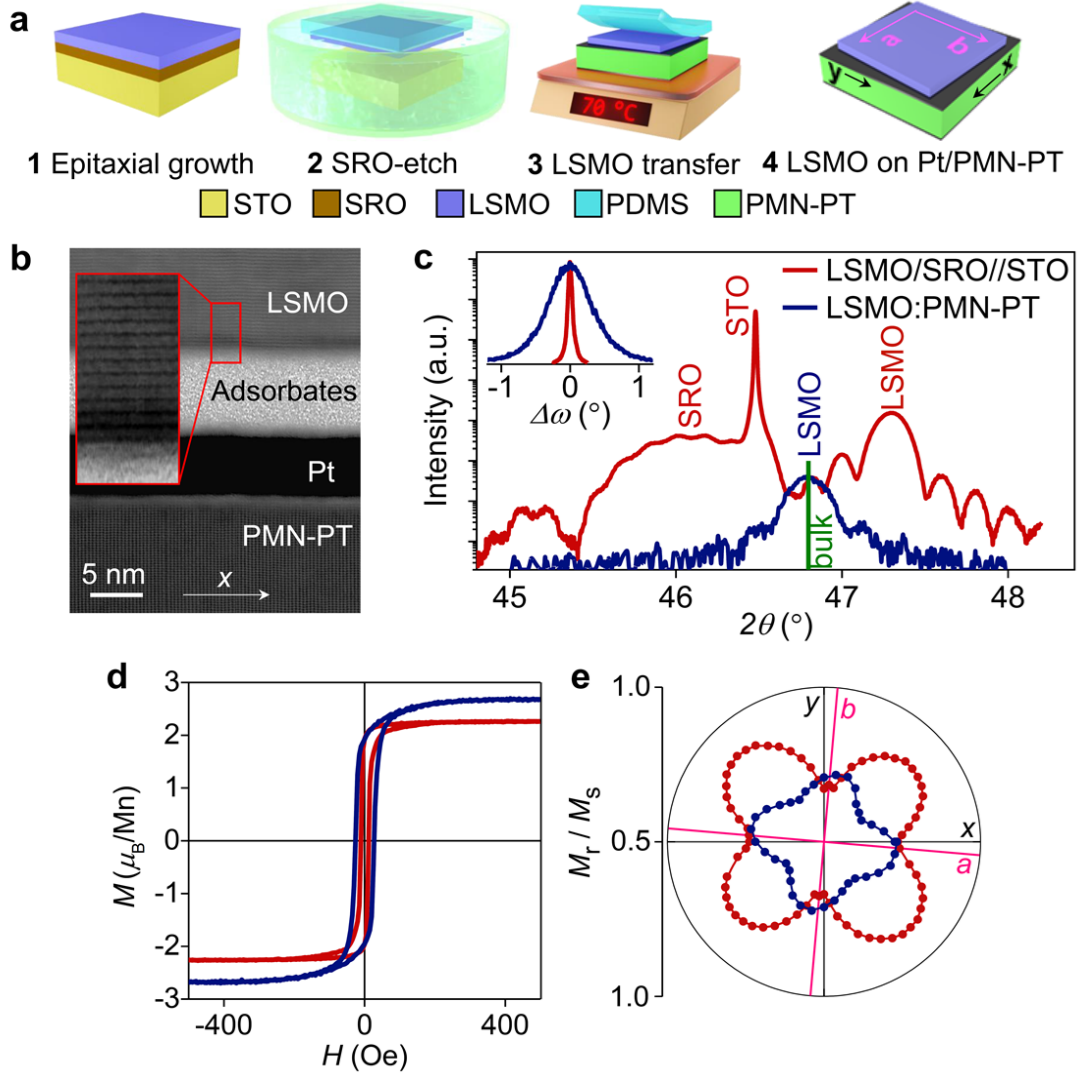

**Figure 1. Transfer of epitaxial LSMO (001)$_{pc}$ to platinized PMN-PT (011)$_{pc}$.** (a) The four-step transfer process in which we approximately aligned the edges of the LSMO film (along $a \parallel [100]_{pc}$ and $b \parallel [010]_{pc}$) with the edges of the slightly larger PMN-PT substrate (along $x \parallel [100]_{pc}$ and $y \parallel [01\bar{1}]_{pc}$). (b) Bright-field cross-sectional STEM image of the interfacial region between LSMO and PMN-PT, looking down the $[01\bar{1}]_{pc}$ zone axis of PMN-PT. The region magnified by ×3.5 confirms in-plane LSMO misalignment. (c-e) Data for electrically virgin LSMO:PMN-PT (blue) and its LSMO/SRO//STO precursor (red). (c) XRD $2\theta$-$\omega$ scans showing $002_{pc}$ reflections. Green vertical line corresponds to the expected $002_{pc}$ reflection for bulk LSMO with pseudocubic lattice parameter[60] 3.881 Å. Inset: $002_{pc}$ rocking curves for LSMO. (d) Magnetization $M$ versus collinear applied field $H$ for one of the two easy axes. (e) Polar plot of in-plane loop squareness $M_r/M_s$, where $M_r$ denotes remanent magnetization and $M_s$ denotes saturation magnetization. Data in (b) for sample C. Data in (c-e) for sample A and its precursor.

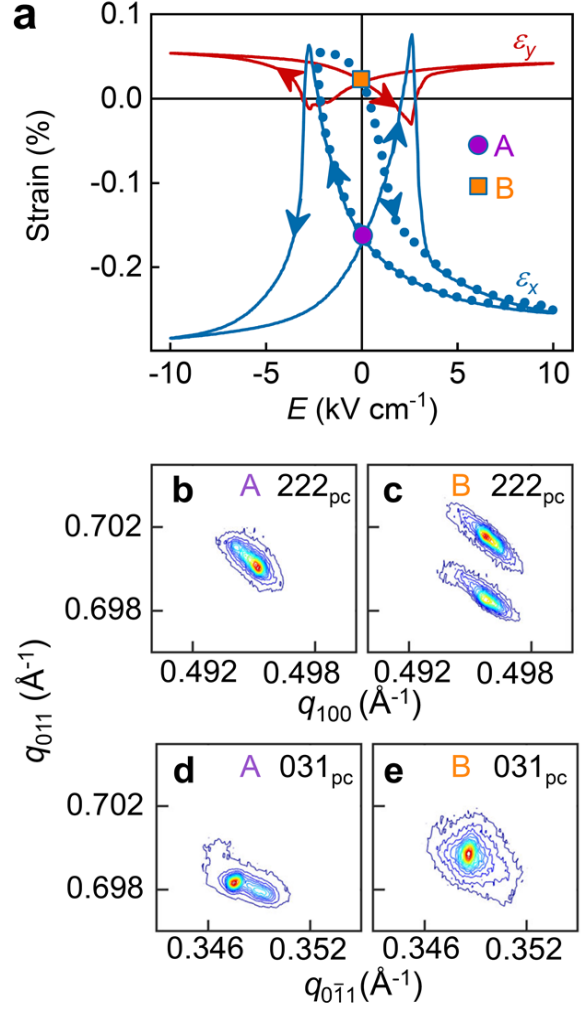

**Figure 2. Electrically driven strain in PMN-PT (011)$_{pc}$.** (a) In-plane strain components $\varepsilon_x$ (blue) and $\varepsilon_y$ (red) versus electric field $E$, for PMN-PT with $x \parallel [100]_{pc}$ and $y \parallel [01\bar{1}]_{pc}$. Remanent states A and B were achieved via the minor loop shown for $\varepsilon_x$ (dotted blue line). (b-e) Reciprocal space maps showing the (b,c) 222$_{pc}$ and (d,e) 031$_{pc}$ reflections for PMN-PT at (b,d) A and (c,e) B. Scattering vector $q = (2/\lambda)\sin(\theta)$ for Bragg angle $\theta$. Data in (a) for annealed PMN-PT from the master substrate used for sample A. Data in (b-e) for sample A.

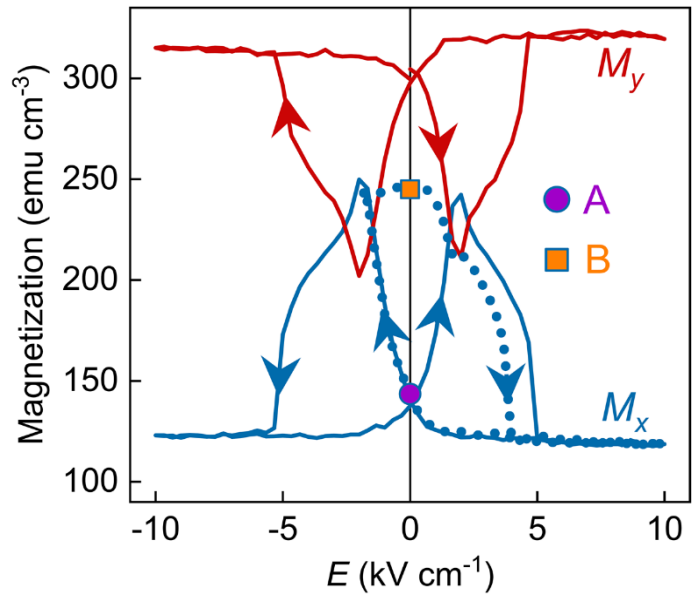

**Figure 3. Strain-mediated electrical control of macroscopic magnetization in LSMO:PMN-PT.** In-plane magnetization components $M_x$ (blue) and $M_y$ (red) versus electric field $E$. Remanent states A and B were achieved via the minor loop shown for $M_x$ (dotted blue line). Data were obtained after applying and removing 1 kOe along the measurement direction, using sample A once it had undergone 30 bipolar electrical cycles (Supplementary Note 7). Similar major loops of $M_x(E)$ were obtained for samples B and C (Supplementary Note 8).

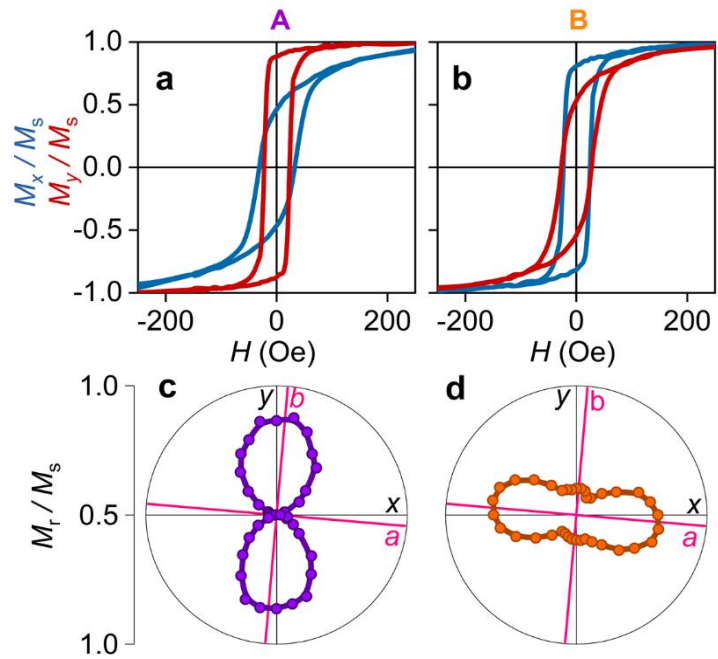

**Figure 4. Electrically controlled magnetic anisotropy in LSMO:PMN-PT.** For remanent states (a,c) A and (b,d) B, we show (a,b) reduced magnetization components $M_x/M_s$ (blue) and $M_y/M_s$ (red) versus collinear applied field $H$, and (c,d) polar plots of loop squareness $M_r/M_s$ derived from plots that include those shown in (a,b). Data for sample A after 30 bipolar electrical cycles (Supplementary Note 7).

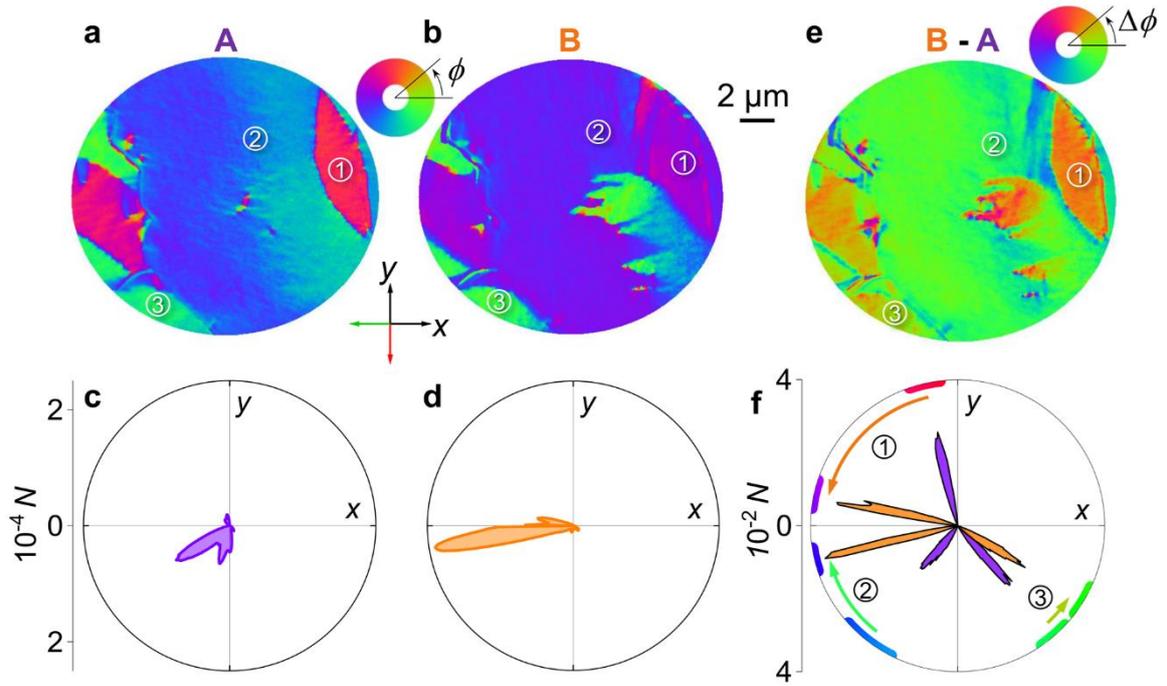

**Figure 5. Electrically induced rotations of magnetic domains in LSMO:PMN-PT.** For remanent states (a,c) A and (b,d) B, we show the number of pixels $N$ in our 20 μm-diameter field of view with in-plane magnetization direction $\phi$ on (a,b) XMCD-PEEM vector maps and (c,d) polar plots. Comparison of (a,b) yields (e) a map of changes in pixel magnetization direction $\Delta\phi$. (f) Polar plot describing 1.3 μm-diameter regions [1-3 in (a,b,e)] at electrically remanent states A (purple lobes) and B (orange lobes). Pixel colours and angular changes are marked on the periphery. Green and red arrows denote the in-plane projections of the grazing-incidence x-ray beam used to obtain (a,b). All data were obtained for sample A following macroscopic magnetoelectric measurements and subsequent thermal demagnetization.

# Supplementary Information
# Large magnetoelectric coupling in multiferroic oxide heterostructures assembled via epitaxial lift-off

D. Pesquera[1,*], E. Khestanova[2], M. Ghidini[3,4,1], S. Zhang[1,5], A. P. Rooney[6], F. Maccherozzi[4], P. Riego[1,7,8], S. Farokhipoor[9], J. Kim[1], X. Moya[1], M. E. Vickers[1], N. A. Stelmashenko[1], S. J. Haigh[6], S. S. Dhesi[4] and N. D. Mathur[1,†]

[1]Department of Materials Science, University of Cambridge, Cambridge, CB3 0FS, UK
[2]ITMO University, Saint Petersburg 197101, Russia
[3]Department of Mathematics, Physics and Computer Science, University of Parma, 43124 Parma, Italy
[4]Diamond Light Source, Chilton, Didcot, Oxfordshire, OX11 0DE, UK
[5]College of Science, National University of Defense Technology, Changsha 410073, China
[6]School of Materials, University of Manchester, Manchester M13 9PL, UK
[7]CIC nanoGUNE, E-20018 Donostia-San Sebastian, Spain
[8]Department of Condensed Matter Physics, University of the Basque Country, UPV/EHU, E-48080 Bilbao, Spain
[9]Zernike Institute for Advanced Materials, University of Groningen, 9747 AG Groningen, The Netherlands

*dpesquera@cantab.net, †ndm12@cam.ac.uk

## Contents

**Note 1. In-plane orientation of PMN-PT substrate and transferred LSMO film**
**Note 2. The interfacial layer between transferred LSMO and platinized PMN-PT**
**Note 3. Surface topography of transferred LSMO and platinized PMN-PT**
**Note 4. Cracks in the LSMO film**
**Note 5. XRD measurements of the transferred LSMO film**
**Note 6. Electrically driven phase transition in PMN-PT**
**Note 7. Creation of the A and B remanent states in LSMO:PMN-PT**
**Note 8. Reproducibility of macroscopic magnetoelectric measurements**
**Note 9. Influence of cracks on the magnetic domain structure**
**Note 10. Simulation of magnetoelectric effects**

## Note 1. In-plane orientation of PMN-PT substrate and transferred LSMO film

In-plane phi-scans around high-intensity reflections (Fig. S1) confirmed a 2-fold symmetry in our PMN-PT $(011)_{pc}$ substrates, and revealed 4-fold symmetry in our transferred LSMO $(001)_{pc}$ films. A comparison of phi-scans reveals that the high-symmetry axes ([010] || $b$ in LSMO, $[01\bar{1}]$ || $y$ in PMN-PT) were misaligned by 5° (Samples A,B) or 20° (Sample C).

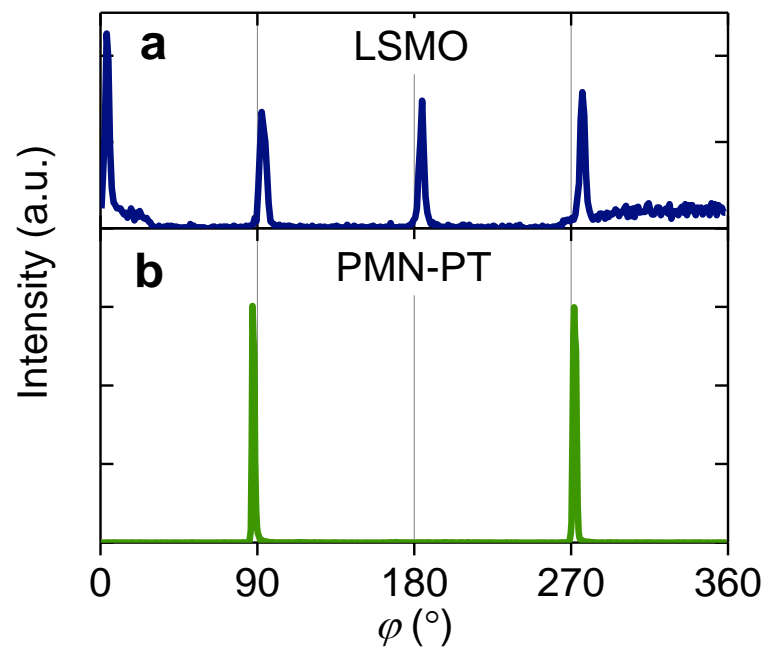

**Figure S1. In-plane phi-scans around high-intensity reflections in LSMO and PMN-PT.** (a) Family of $011_{pc}$ reflections for LSMO. (b) Family of $013_{pc}$ reflections for PMN-PT. The misorientation between the crystal axes of two materials is 5°. Data for Sample A.

**Note 2. The interfacial layer between transferred LSMO and platinized PMN-PT**

Cross-sectional scanning transmission electron microscopy (STEM) revealed the presence of a 10 nm-thick amorphous layer between the LSMO film and top electrode of the platinized PMN-PT substrate (Fig. S2a). This amorphous layer, which mediates the strain-mediated magnetoelectric effects that we observed, was found to be flat and uniform over several microns in our STEM lamella. The maximum thickness variation was 2 nm, and we may infer that contaminants did not self-assemble into large pockets[S1].

Elemental analysis revealed that the amorphous layer contained Si, O and accumulation of C at both interfaces (Fig. S2b). We attribute these residuals to the partial degradation of the PDMS membrane during the $SrRuO_3$ etch (no residuals of Sr or Ru are apparent). Carbon may also be present near the Pt electrode due to adsorbed hydrocarbons, which typically become trapped between substrates and transferred two-dimensional crystals[S2,S3].

Electron energy loss spectroscopy (EELS) images obtained at the Mn $L_3/L_2$ edges (not shown) did not identify a change of Mn oxidation state in the LSMO film near its interface with the amorphous layer. This implies van-der-Waals bonding rather than chemical reaction, as suggested for LSMO transferred to Si substrates[S4].

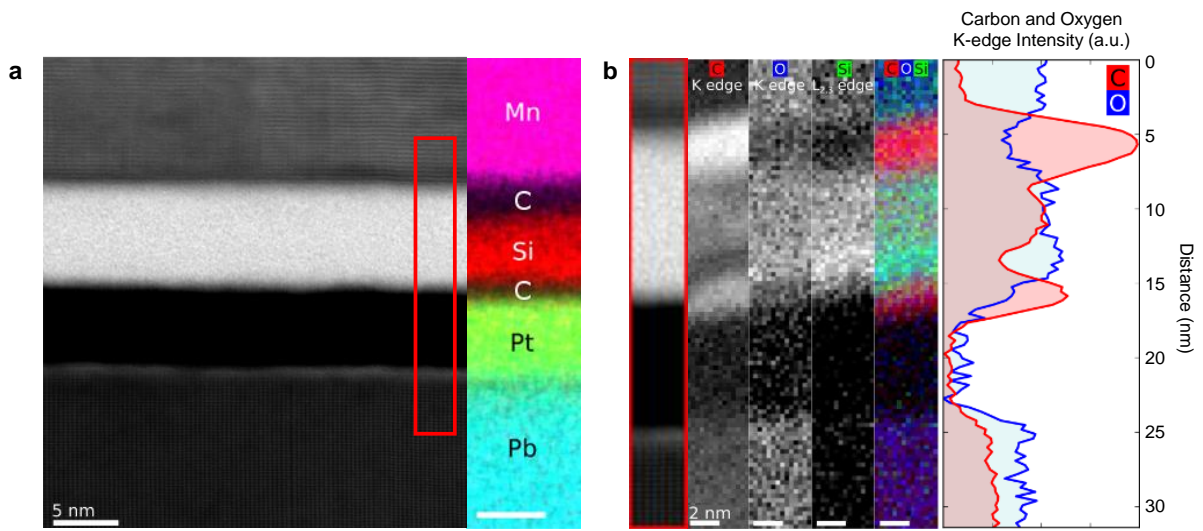

**Figure S2. Detail showing the amorphous layer between LSMO and the platinized PMN-PT substrate.** (a) Bright-field cross-sectional STEM image for LSMO:PMN-PT, and energy dispersive x-ray (EDX) image of the adjacent region showing selected elements in each layer. Scale bars in (a) are 5 nm. (b) The region highlighted red in (a) is magnified at left. The other panels in (b) show EELS data for this area, namely images obtained at the as-specified carbon, oxygen and silicon edges (presented separately and together); and the K-edge intensity distribution for C and O (normalised counts do not indicate relative composition). Scale bars in (b) are 2 nm. All data for sample C.

**Note 3. Surface topography of transferred LSMO and platinized PMN-PT**

Atomic force microscopy images of the transferred LSMO film (Fig. S3a) and an exposed region of the platinized PMN-PT substrate (Fig. S3b) are both dominated by topographical features that arise from ferroelectric domains spanning roughly 1-5 μm.

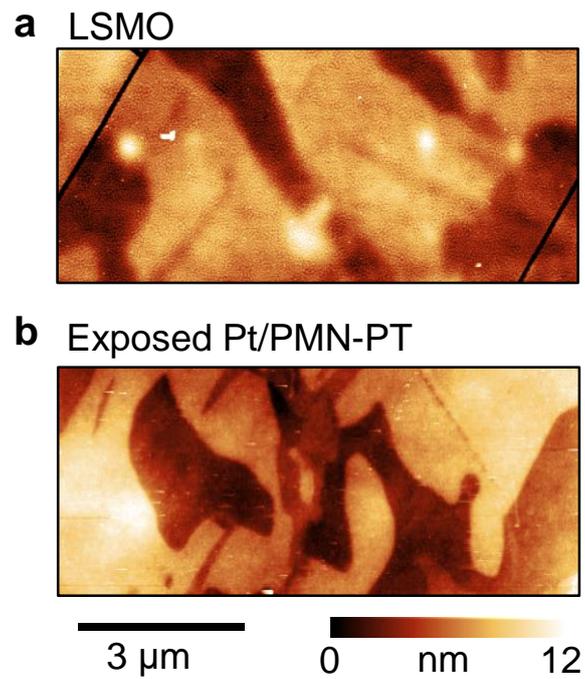

**Figure S3. Surface topography for LSMO:PMN-PT.** Atomic force microscopy (AFM) images of (a) transferred LSMO and (b) the exposed Pt/PMN-PT. Data from sample B.

## Note 4. Cracks in the LSMO film

We observed cracks in the LSMO film after transfer to the PDMS membrane (inset, Fig. S4), likely due to epitaxial stress relief after release from the STO substrate. It therefore follows that we observed cracks in the LSMO film after it had been subsequently transferred to the platinized PMN-PT substrate (Fig. S4).

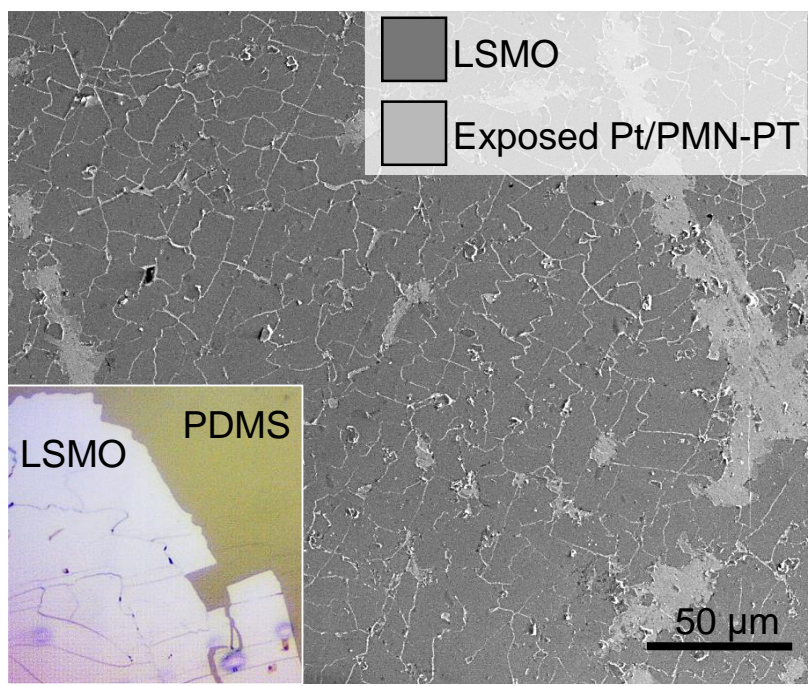

**Figure S4. Cracks in the LSMO film during and after transfer.** Scanning Electron Microscopy (SEM) image of LSMO:PMN-PT showing the transferred LSMO film (dark) and exposed regions of the platinized PMN-PT substrate (bright). Inset: optical microscopy image of the same LSMO film on the PDMS membrane before transfer to the platinized PMN-PT (no scale recorded). Data for Sample C (main image) and precursor LSMO (inset).

**Note 5. XRD measurements of the transferred LSMO film**

Our transferred LSMO film displayed split x-ray reflections (Fig. S5a,b), permitting us to infer that it comprised twins of orthorhombic LSMO with different in-plane lattice parameters.

All four reflections that we show in Fig. S5 undergo small shifts on electrically switching between the remanent states that we label A and B. These shifts are only just resolved, which we infer to be a consequence of measuring small twins on a topography determined by the underlying ferroelectric domains.

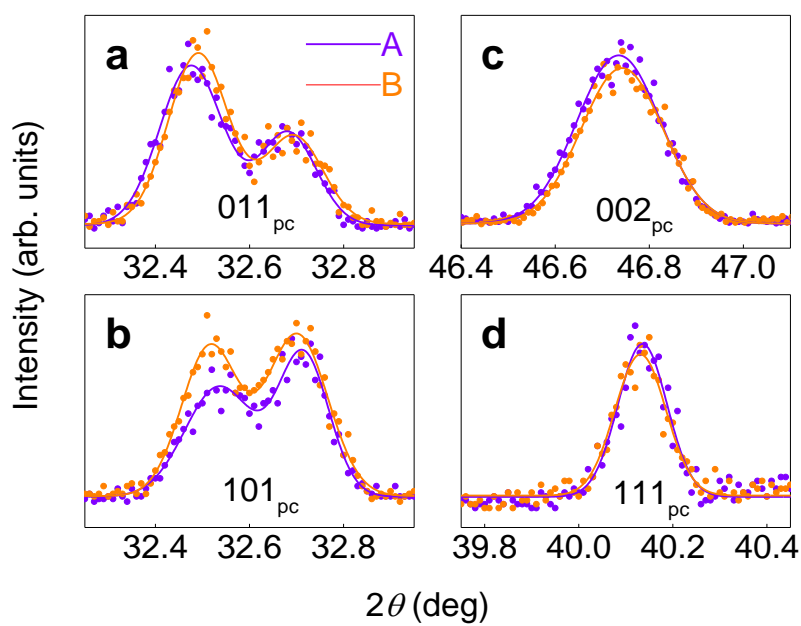

**Figure S5. Structure of the LSMO film at remanent states A and B in LSMO:PMN-PT.** (a-d) X-ray diffraction $2\theta$-$\omega$ scans showing four LSMO reflections in remanent states A (purple) and B (orange). Lines are fits to data. Data for Sample A.

## Note 6. Electrically driven phase transition in PMN-PT

Here we show that electrically induced switching of electrical polarization (Fig. S6a) and macroscopic strain (Fig. S6b,c) are associated with the interconversion of orthorhombic (O) and rhombohedral (R) phases. We identify this interconversion from reciprocal space maps (Fig. S6d,e) for the $(222)_{pc}$ and $(031)_{pc}$ lattice planes (Fig. S6f), whose asymmetric x-ray reflections undergo complementary splitting (Fig. S6g,h) due to the presence of domains (Fig. S6i,j).

Specifically, our reciprocal space maps for the $222_{pc}$ (Fig. S6d) and $031_{pc}$ (Fig. S6e) reflections reveal that:

- There was essentially a single broad peak after thermal depolarization.
- On applying and removing -10 kV cm$^{-1}$, there was a single $222_{pc}$ reflection and a split $031_{pc}$ reflection, implying the formation of $O_{1,2}$ domains.
- On approaching the peak-strain at 2.67 kV cm$^{-1}$, the single $222_{pc}$ reflection observed after poling was replaced by a split peak, and the split $031_{pc}$ reflection observed after poling was replaced by a single peak, implying the formation of $R_{1,2}$ domains. The concomitant formation of minority $R_{3,4}$ domains with in-plane polarizations[S5] is evidenced by the creation and subsequent destruction of an additional $031_{pc}$ reflection with low intensity.
- At 3.33 kV cm$^{-1}$, the $O_{1,2}$ domains began to reappear.
- At higher fields, the switching current approached zero, the strain approached saturation, the $O_{1,2}$ domains were re-established and the R phase was eliminated.

The two strain states that we achieved at electrical remanence[S6] (Fig. 2b in main paper) are therefore identified with the O and R phases as follows:

- The A state obtained after removing a saturating field is identified with the O phase observed here after likewise poling.
- The B state obtained after reducing a saturating field through zero to the coercive field of opposite sign, and then removing this coercive field, is identified with the R phase observed here near the switching field.

Note that states A and B were obtained after positive poling (Fig. 2b in main paper), while the corresponding states identified here were obtained after negative poling (Fig. S6d,e), such that permitted directions of polarization (red arrows, Fig. S6i,j) are reversed.

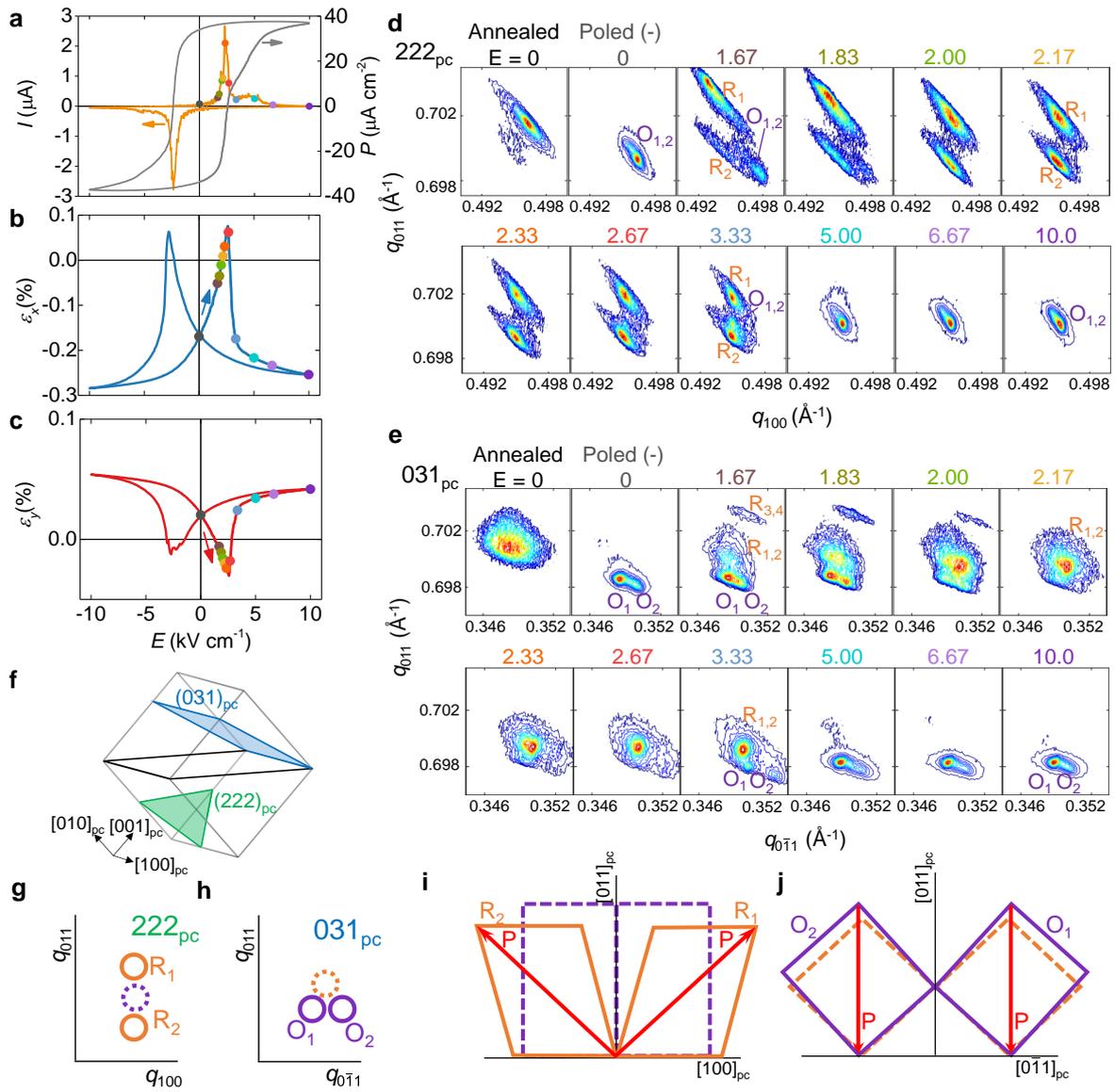

**Figure S6. Electrically driven phase transition in PMN-PT.** (a-c) For major loops of electric field $E$, we show (a) polarization $P$ (grey) deduced from switching current $I$ (orange), and in-plane strains (b) $\varepsilon_x$ and (c) $\varepsilon_y$. Coloured points identify coloured kV cm$^{-1}$ values of $E$ in (d,e), where we show reciprocal space maps around the (d) $222_{pc}$ and (e) $031_{pc}$ reflections obtained first at zero-field after a rejuvenating anneal in air at 150 °C, then at zero-field after negative poling, and then at increasing values of $E$. Scattering vector $q = (2/\lambda)\sin(\theta)$ for Bragg angle $\theta$. Data in (a-c) acquired simultaneously using PMN-PT from the same master as sample A. Data in (a-e) for Sample A. (f) Pseudocubic unit cell showing the $(011)_{pc}$ surface (two bold black lines) and the $(222)_{pc}$ and $(031)_{pc}$ lattice planes. (e-j) For the R-phase (orange) and O-phase (purple), schematics of the reciprocal space maps show the (g) $222_{pc}$ and (h) $031_{pc}$ reflections, which are split (solid outlines) or unsplit (dashed outlines) according to whether (i,j) the projections of the unit cell over the corresponding real-space planes present twinning (solid outlines) or not (dashed outlines). Red arrows in (i,j) denote permitted directions of polarization.

# Note 7. Creation of the A and B remanent states in LSMO:PMN-PT

The four-fold anisotropy in LSMO after transfer (Fig. 1e of the main paper, reproduced in Fig. S8a) was modified by 30 bipolar cycles that each comprised major or minor loops of the type used to measure macroscopic magnetoelectric effects (Fig. 3 of the main paper). The inclusion of minor loops permitted access to remanent states A and B, whose evolution is shown in Fig. S8b-d.

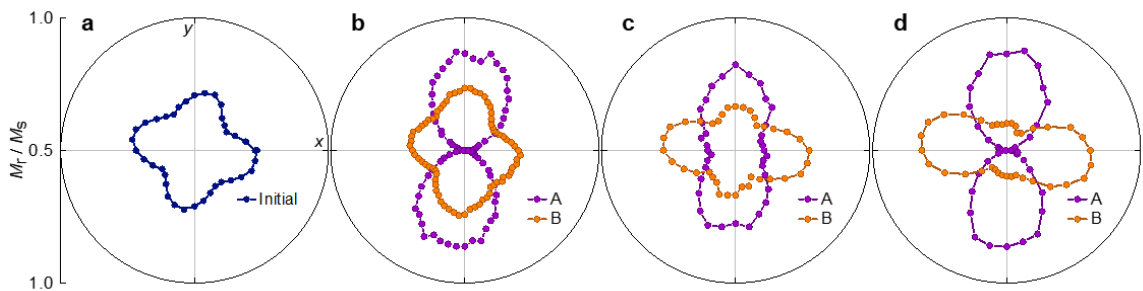

**Figure S8. Creation of the A and B remanent states in LSMO:PMN-PT.** (a-d) Polar plots of loop squareness $M_r/M_s$ based on data obtained (a) for the virgin state, and then after (b) 2, (c) 5 and (d) 30 bipolar cycles (that each comprised either a major loop accessing remanent state A, or a minor loop accessing remanent states A and B). Magnetic remanence is denoted $M_r$, saturation magnetization is denoted $M_s$. Data in (a) matches data in Fig. 1e of the main paper. Data in (d) matches data in Fig. 4c,d of the main paper.

## Note 8. Reproducibility of macroscopic magnetoelectric measurements

Major loops of $M_x(E)$ for sample A (Fig. 3 in the main paper) and two similar samples are similar (Fig. S7).

Fig. S7 also shows that magnetoelectric effects are completely suppressed by a saturating magnetic field. This confirms our expectation[S7] that there are no strain-driven changes in the magnitude of the local magnetic moment, consistent with the magnetic rotations that we observed (Fig. 5 in the main paper).

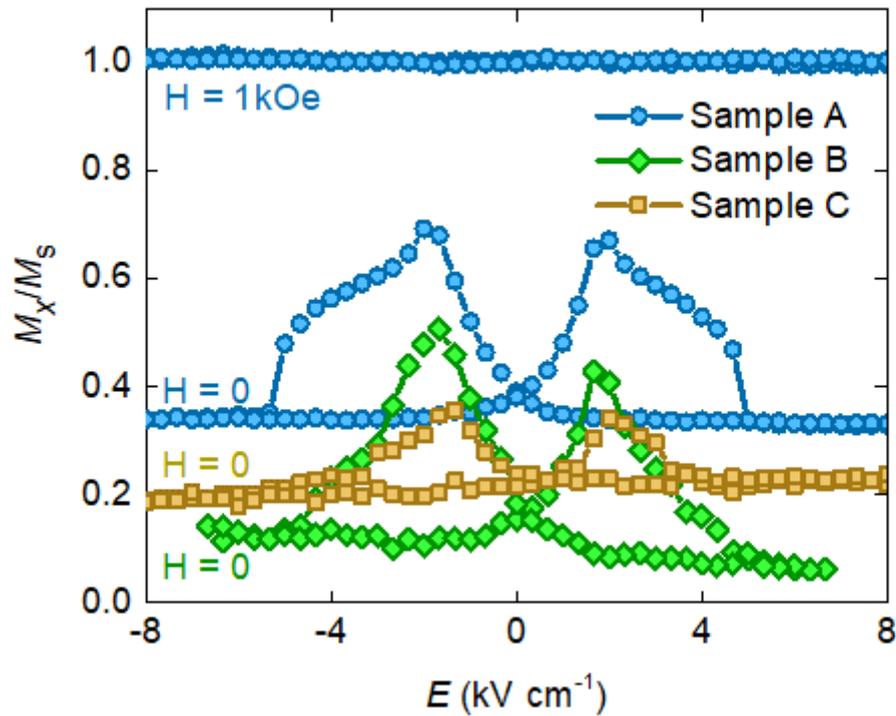

**Figure S7. Macroscopic magnetoelectric effects in LSMO:PMN-PT.** In-plane magnetization component $M_x$ normalized by saturation magnetization $M_s$ versus electric field $E$. The lower three plots were measured for Samples A-C in zero magnetic field $H$, after applying and removing $H$ = 1 kOe along $x$. The uppermost plot was measured for Sample A with $H$ = 1 kOe along $x$.

## Note 9. Influence of cracks on the magnetic domain structure

In three regions of the LSMO:PMN-PT sample, cracks (left panels in Fig. S9) were found to coincide with a subset of magnetic domain perimeters (right panels in Fig. S9).

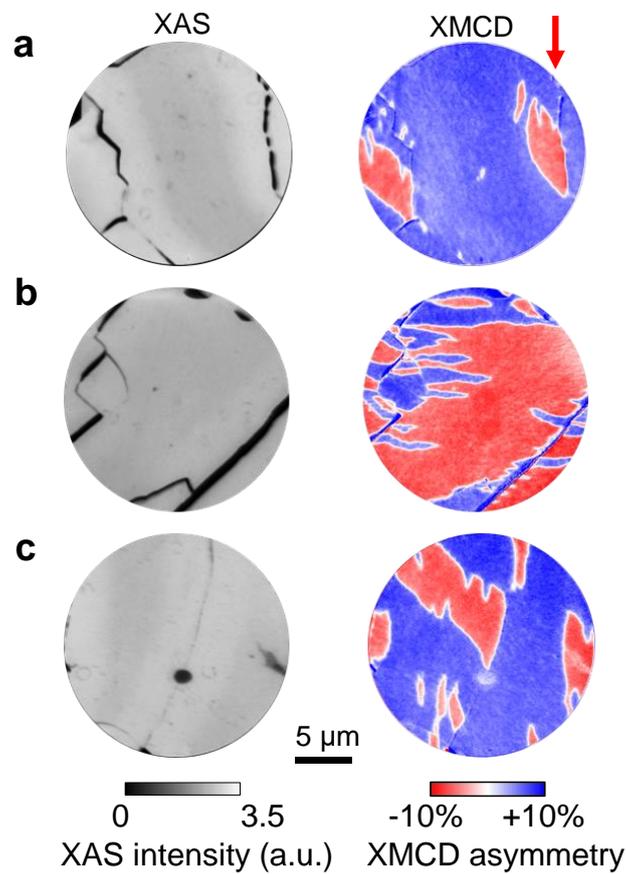

**Figure S9. Correlation between cracks and magnetic domains in LSMO:PMN-PT.** Photoemission electron microscopy (PEEM) images for three different areas of the sample, with contrast from (left) x-ray absorption spectroscopy (XAS) and (right) x-ray magnetic circular dichroism (XMCD). Red arrow denotes the in-plane projection of the grazing-incidence x-ray beam. Data for Sample A.

## Note 10. Simulation of magnetoelectric effects

First, we will present a free energy density that describes the magnetoelectric effects in our tranferred LSMO film, where the observed inhomogeneity is attributed to an invariant uniaxial stress anisotropy arising from the transfer. Then we will use this free energy density to simulate macroscopic and microscopic magnetoelectric effects, for comparison with our experimental observations.

**Free energy density**

The inhomogeneity that we observed in our XMCD-PEEM vector maps (Fig. 5 in the main paper) evidences different types of region in our transferred LSMO film. We will arbitrarily consider there to be nine types of region, and we will neglect exchange coupling between adjacent regions. The free energy density $F$ for the LSMO film may then be written as:

$$F(E) = \sum_{i=1}^{9} F_i(\phi_i, E),$$

where $F_i$ denotes the total free energy density for all distributed regions of the $i^{th}$ type, where the local magnetization in regions of the $i^{th}$ type adopts direction $\phi_i$, where the magnitude of the local magnetization is equal to the saturation magnetization $M_s = 425$ emu cm$^{-3}$ of the transferred LSMO film (blue data, Fig. 1d in the main paper), and where $E$ denotes the electric field applied across the PMN-PT substrate. The function $F_i(\phi_i, E)$ is given by:

$$F_i(\phi_i, E) = -K_c \sin^2(\phi_i - \alpha)\cos^2(\phi_i - \alpha) - K_s(\varepsilon_{\text{eff}}(E))\cos^2(\phi_i - \beta) - K_u \cos^2(\phi_i - \varphi_i),$$

where each term on the right is described below. Positive angles ($\phi_i$, $\alpha$, $\beta$ and $\varphi_i$) imply anticlockwise rotations with respect to the $x$ direction in PMN-PT as viewed from the LSMO film.

<u>Magnetocrystalline anisotropy</u> The term $-K_c \sin^2(\phi_i - \alpha)\cos^2(\phi_i - \alpha)$ is common to all nine types of region in the LSMO film, and describes the four-fold magnetocrystalline anisotropy (ref. 45 in the main paper), for which $\alpha = 40°$ is one of the hard directions (Fig. 1e in the main paper). The anisotropy constant $K_c = H_a M_s/2 = 3.5$ kJ m$^{-3}$ was estimated from the values of anisotropy field $H_a = 165$ Oe and saturation magnetization $M_s = 425$ emu cm$^{-3}$ that we measured for the transferred LSMO film (blue data, Fig. 1d in the main paper). Fig. S10a shows a polar plot of the magnetocrystalline anisotropy density for the LSMO film.

Piezostress anisotropy The term $-K_s(\varepsilon_{\text{eff}}(E))\cos^2(\phi_i - \beta)$ is common to all nine types of region in the LSMO film, and describes the uniaxial stress anisotropy due to piezoelectric strain from the substrate. For this term, $\beta = 0°$ is the hard direction for the negative value of anisotropy constant $K_s(\varepsilon_{\text{eff}}(E))$ in state A, and $\beta = 90°$ is the hard direction for the positive value of anisotropy constant $K_s(\varepsilon_{\text{eff}}(E))$ in state B. The anisotropy constant $K_s(\varepsilon_{\text{eff}}(E)) = \frac{3}{2}\lambda E_Y \varepsilon_{\text{eff}}(E)$ changes sign due to electrically driven changes in the effective strain[S8] $\varepsilon_{\text{eff}}(E) = \eta[\varepsilon_x(E) - \varepsilon_y(E)]/(1 + \nu)$ experienced by the LSMO film, where:

- $\varepsilon_x(E)$ and $\varepsilon_y(E)$ are the electric-field dependent strains transmitted by the PMN-PT substrate (Fig. 2a in the main paper);
- strain-transfer coefficient $\eta = 40\%$ was found to be reasonable by comparing our results for $M_x(E)$ and $M_y(E)$ (Fig. S11) with our macroscopic measurements (Fig. 3 in the main paper);
- the LSMO Poisson's ratio $\nu = 0.33$ was calculated from the strain in our epitaxial LSMO film prior to transfer (the XRD data in Fig. 1c in the main paper implies an out-of-plane pseudocubic lattice parameter of 3.841 Å, the SrTiO$_3$ substrate implies an in-plane pseudocubic lattice parameter of 3.905 Å, and the bulk lattice parameter of LSMO was taken to be[S9] 3.873 Å);
- we assume $\varepsilon_z = 0$;
- LSMO magnetostriction[S10] $\lambda = 3 \times 10^{-5}$;
- LSMO Young's modulus[S11] $E_Y = 100$ GPa.

Fig. S10b shows a polar plot of the electrically controlled stress anisotropy density for the LSMO film.

Constant stress anisotropy The term $-K_u \cos^2(\phi_i - \varphi_i)$ describes a uniaxial stress anisotropy, which we assume to arise in the $i^{\text{th}}$ type of region due to stress arising from the transfer of the LSMO film (we set $K_u = K_c$). The nine types of region are characterized by nine equally separated directions $\varphi_i = \frac{\pi i}{9}$. Each of these directions describes an easy axis for $-K_u \cos^2(\phi_i - \varphi_i)$, and collectively the nine directions average to zero (consistent with our macroscopic measurements of strain and magnetic anisotropy). Fig. S10c shows polar plots of the constant stress anisotropy for each type of region. The free energy density $F_i(\phi_i,E)$ for each type of region is shown in Fig. S10d for state A, and in Fig. S10e for state B.

The free energy density $F(E)$ for the LSMO film displays a two-fold anisotropy (Fig. 10f), with the easy axis along *y* (state A) or *x* (state B), as observed experimentally (Fig. 3 in the main paper).

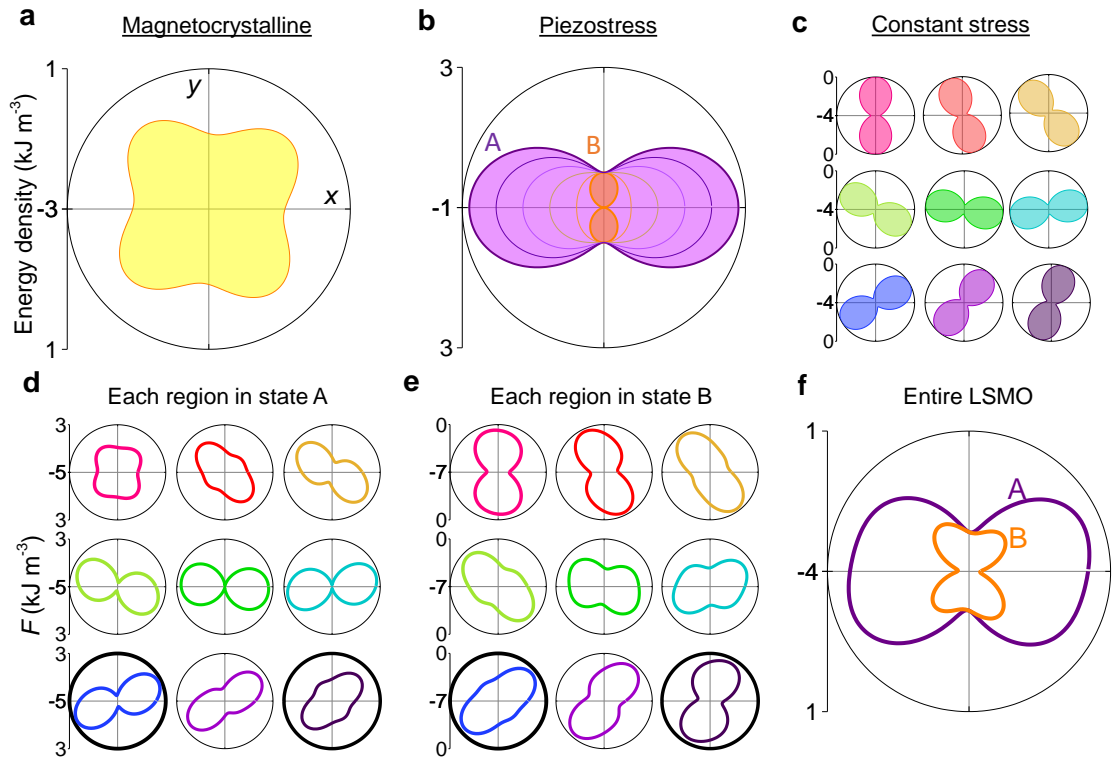

**Figure S10. Polar plots of energy density.** (a) Polar plot of the magnetocrystalline anisotropy density, which is common to all nine types of region in the LSMO film. (b) Polar plot of the electrically controlled stress anisotropy density, which is common to all nine types of region in the LSMO film, and differs for states A and B. (c) Polar plots of the constant stress anisotropy density for the nine types of region in our LSMO film that encode the observed inhomogeneity. (d,e) Polar plots of the free energy density $F_i(\phi_i,E)$ for all nine regions when the sample adopts (d) state A and (e) state B (data obtained by summing the plots in (a-c), the four plots with bold outlines are shown in the main paper as Fig. 5g,h). (f) Polar plot of the free energy density $F(E)$ for the entire LSMO film, when the sample adopts states A and B (data obtained by summing the plots in (d) and (e), respectively).

**Simulation of macroscopic magnetoelectric effects**

The free energy density $F(E)$ for the LSMO film was used to simulate plots of $M_x(E)$ and $M_y(E)$ at magnetic remanence, for comparison with our macroscopic magnetoelectric measurements (Fig. 3 in the main paper).

To account for the initial application and removal of the saturating magnetic field along the measurement axis, we added the magnetostatic energy density $-M_s H \cos(\phi_i - \gamma)$ to the free energy density $F_i(\phi_i, E)$ for each type of region. Here, $H = 1000$ Oe denotes the magnitude of the saturating field that we used in our experiments, $\gamma$ denotes the measurement direction ($\gamma = 0°$ for $x$ and $\gamma = 90°$ for $y$), and $M_s = 425$ emu cm$^{-3}$ (see earlier).

After thus obtaining the nine values of $\phi_i$ that describe the state of the film at magnetic remanence, we used the following three steps to identify $M_x(E)$ and $M_y(E)$ (Fig. S11). First, we used our experimentally obtained plots of $\varepsilon_x(E)$ and $\varepsilon_y(E)$ (Fig. 2a in the main paper) to obtain $\varepsilon_{\text{eff}}(E)$. Second, we used $\varepsilon_{\text{eff}}(E)$ to minimize $F_i(\phi_i, E)$ for each type of region, and thus obtain the nine magnetization directions $\phi_i$ for every tenth experimental value of $E$. Third, we summed the projections of the nine magnetization vectors along $x$ to obtain $M_x(E)$, and along $y$ to obtain $M_y(E)$.

Our plots of $M_x(E)$ and $M_y(E)$ (Fig. S11) are similar to our experimental results (Fig. 3 in the main paper). The constant stress anisotropy term is responsible for rendering the peaks in $M_x(E)$ and the troughs in $M_y(E)$ as broad as their experimental counterparts. Moreover, the constant stress anisotropy term renders the plots of $M_x(E)$ and $M_y(E)$ repeatable on further electric-field cycling.

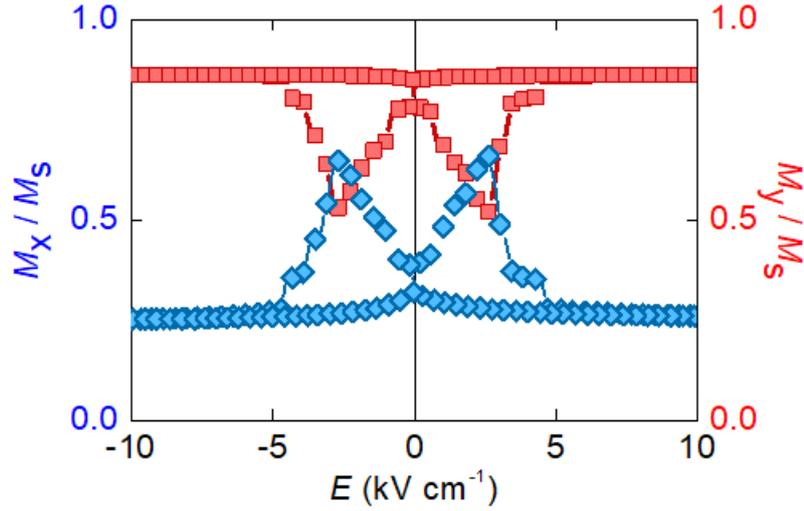

**Figure S11. Calculated magnetoelectric effects.** Predicted plots of $M_x(E)$ and $M_y(E)$, normalized by saturation magnetization $M_s$.

**Simulation of microscopic magnetoelectric effects**

The free energy density $F(E)$ for the LSMO film was used to simulate electrically driven magnetic domain rotations in zero-magnetic field, for comparison with our corresponding microscopic observations based on XMCD-PEEM vector maps (Fig. 5a-f in the main paper).

To identify the demagnetized starting state at $E = 0$, we set the initial directions of magnetization in each type of region at random, with the constraint that all nine directions be equally spaced in angle. We then minimised $F_i(\phi_i, E)$ for each type of region, in order to obtain nine values of $\phi_i$ that no longer averaged exactly to zero.

After thus obtaining the nine values of $\phi_i$ that describe the demagnetized starting state at $E = 0$, we used the following three steps to identify the electrically driven angular changes $\Delta\phi_i$ between states A and B (Table S1, Fig. S12). First, we used our experimentally obtained plots of $\varepsilon_x(E)$ and $\varepsilon_y(E)$ (Fig. 2a in the main paper) to obtain $\varepsilon_{\text{eff}}(E)$. Second, at every tenth experimental value of $E$, we used $\varepsilon_{\text{eff}}(E)$ to minimize $F_i(\phi_i, E)$ for each type of region and thus obtain nine magnetization directions $\phi_i$. Third, we identified the nine values of $\phi_i$ for states A and B in the electrical cycle.

The resulting values of $\Delta\phi_i$ reproduce key features of our experimental observations because they are widely distributed in magnitude (from a few degrees up to 62°), and because some rotation angles calculated here (Regions of type 7, 9 and 3, Table S1) are similar to those observed experimentally (Regions 1-3, Fig. 5f in the main paper).

| Region of type $i$ | $\varphi_i$ (°) | $\phi_i$ in state A (°) | $\phi_i$ in state B (°) | $\Delta\phi_i$ (°) | Colour |
|---|---|---|---|---|---|
| 1 | 0 | 176 | 178 | 2 | ■ |
| 2 | 20 | -111 | -173 | -62 | ■ |
| 3 | 40 | -109 | -161 | -52 | ■ |
| 4 | 60 | -102 | -112 | -10 | ■ |
| 5 | 80 | -95 | -99 | -4 | ■ |
| 6 | 100 | -88 | -87 | 1 | ■ |
| 7 | 120 | -81 | -68 | 13 | ■ |
| 8 | 140 | -74 | -21 | 53 | ■ |
| 9 | 160 | 108 | 169 | 61 | ■ |

**Table S1. Calculated magnetization directions.** For the nine types of region defined by the direction $\varphi_i$ of the constant uniaxial stress anisotropy, we give values of $\phi_i$ for states A and B. These values and the corresponding changes $\Delta\phi_i$ are presented in Fig. S12 using the colour code defined here.

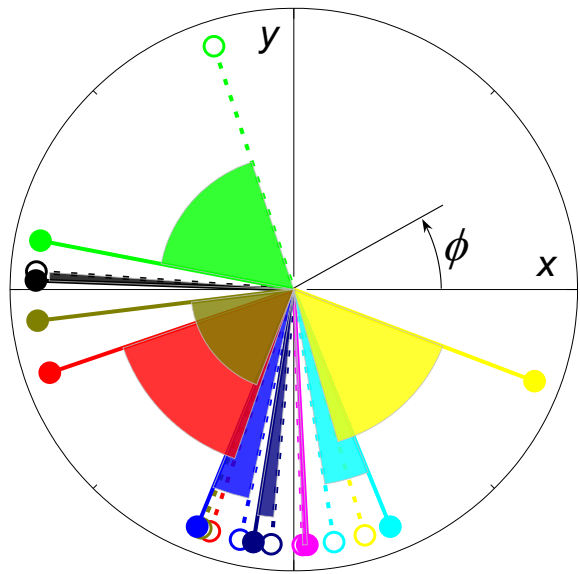

**Figure S12. Calculated magnetization directions.** Visual presentation of the data in Table S1. Values of $\phi_i$ in State A (empty circles) and State B (filled circles) differ by $\Delta\phi_i$ (shaded sectors).